\documentclass[twocolumn,prd]{revtex4}
\usepackage{graphicx}

\IfFileExists{srcltx.sty}{\usepackage[active]{srcltx}}

\begin{document}
\title{Quantum cosmology of the brane universe.}
\author{A.Boyarsky$^1$, A.Neronov$^2$, I.Tkachev$^3$}
\affiliation{$^1$ Ecole Polytechnique F\'ed\'erale de
    Lausanne, BSP 720, CH-1015, Lausanne, Switzerland\\
$^2$ ISDC,CH-1290 Versoix
Switzerland , Switzerland,\\
$^3$ Theory department, CERN, 1211 Geneve 23, Switzerland}
\begin{abstract}
  We canonically quantize the dynamics of the brane universe embedded
  into the five-dimensional Schwarzschild-anti-deSitter bulk
  space-time. We show that in the brane-world settings the formulation
  of the quantum cosmology, including the problem of initial
  conditions, is conceptually more simple than in the 3+1-dimensional
  case. The Wheeler-deWitt equation is a finite-difference equation.
  It is exactly solvable in the case of a flat universe and we find the
  ground state of the system. The closed brane universe can be created
  as a result of decay of the bulk black hole.
\end{abstract}
\maketitle

{\it Introduction.}
Quantum effects almost certainly played crucial role in the early universe
evolution and in the process of universe creation. Understanding and study of
quantum cosmology is important not only from the conceptual point of view,
but, hopefully, may provide us with constraints on possible topology of the
universe and initial conditions for the inflationary stage
\cite{mini,tunneling,Linde:2004nz}. Appropriate theoretical frameworks which
would incorporate all quantum gravitational effects are yet to be constructed,
however.

String theory, eventually, may provide the consistent approach to the quantum
cosmology realm, but the formulation of the string theory on a non-trivial and
significantly Lorentzian space time is very complicated and unsolved task (
see for example~\cite{string} and references therein).  That is why the
approaches based on canonical quantization of the Einstein gravity \cite{wdw}
still prove to be more successful in addressing the problems of quantum
cosmology.  Here one has to adopt a modest approach and restrict consideration
to quantum phenomena below the Plank energy scale. Quantizing the universe as
a whole one has further resort to the ``mini-superspace'' modeling
\cite{mini,tunneling,qg,Linde:2004nz} in order to get to definite final
results (for a recent interesting development see, however, Ref.~\cite{jan}
where effective action for the scale factor was derived integrating out other
gravitational degrees of freedom using numerical simulations).

Even then, within the ``mini-superspace'' approach, many conceptual and     
technical problems remain, such as the problem of ascribing physical meaning
to the wave function of the universe~\cite{qg}.  Other important issues are
the choice of the boundary conditions which one imposes at the big-bang point
(e.g. ``no-boundary'' \cite{qg}, ``tunneling'' \cite{tunneling}, etc.) and
the problem of unboundedness of the gravitational action (see
e.g. \cite{vilenkin_rev}).

In the present paper we pursue the viewpoint that the presence of extra
dimensions can resolve or relax some of these problems.  Indeed, in the brane
world scenario \cite{Rubakov:1983bb,rs}, the problem of quantum cosmology
(i.e.  quantization of gravitational degrees of freedom) is replaced by a much
better defined problem of quantum mechanics of the brane (matter degrees of
freedom) which moves in the bulk space-time. This has several important
consequences. First, one may hope that probabilistic interpretation, initial
and boundary conditions, "tunneling", "scattering" and ``ground'' states of
the Universe become better defined.  
Second, one can escape, to some extent,
solving the problems of quantum gravity. Indeed, the big bang point, i.e. the
point of vanishing brane size, can be unreachable due to quantum uncertainty.
Thus, quantization of matter in a self-consistently calculated ``external''
gravitational field can be sufficient.

The conceptual simplicity of the brane quantum cosmology does not imply its
``technical'' simplicity: one has to take into account self-consistently the
interaction of the brane with the bulk both on classical and quantum
levels. Here we can benefit capitalizing on the fact, that the dynamics of
(3+1)-dimensional brane embedded in (4+1)-dimensional bulk, is very similar to
the dynamics of self-gravitating shells in conventional (3+1)-dimensional
General Relativity, which was studied extensively both at classical
\cite{shell1,shell1a} and quantum levels \cite{shell2,shell3,prd57}.  In the
present paper we generalize the formalism developed in~\cite{prd57} to the
case of the (3+1) dimensional brane universe embedded into (4+1)-dimensional
bulk.

We may hope that some results found in frameworks of brane quantum cosmology
may hold even if the universe is (3+1)-dimensional. In particular, the
distinctive feature of quantum mechanics of branes is that the differential
Schroedinger (or ``Wheeler-deWitt'') equation for the wave function is
replaced by a finite-difference equation \cite{shell2,prd57}. This may
be a general property of ``true'' quantum cosmology. Note in this respect that
finite-difference equations for the wave function of the universe appear also
in the frameworks of loop-quantum gravity \cite{Bojowald}.

{\it Hamiltonian description of the classical motion of a gravitating brane.}
We construct the Hamiltonian formalism which describes the motion of a
self-gravitating thin shell of matter starting from the action of
(4+1)-dimensional Einstein gravity with bulk cosmological constant. The brane
part of the action contains the term proportional to the brane tension
$\lambda$ and the term which describes (in the simplest case) dust-like matter
on the brane with the mass $\mu$ per unit co-moving volume.  The total action
of the system is
\begin{eqnarray}
\label{eq:action}
S&=&\frac{1}{4l_{\rm Pl}^3}\int_{\rm bulk} \sqrt{g}\left[\Lambda+
  ^{(4)}{\cal R}+({\rm Tr} {\cal K})^2-{\rm Tr}
{\cal K}^2\right]\nonumber\\
&&-8\pi\mu\int_{\rm brane}d\tau-\lambda\int_{\rm brane}\sqrt{-\hat g}d\tau 
d^3\hat x \; ,
\end{eqnarray}
where $l_{\rm Pl}^{-1}$ is the (4+1)-dimensional Planck mass, $\hat g$ is the
induced metric on the brane, $\tau$ is the proper time of co-moving observers
in the brane universe, $\Lambda$ is the bulk cosmological constant and
$^{(4)}{\cal R},{\cal K}_{AB}$ are the 4-dimensional Ricci scalar and the
external curvature of the spatial section of (4+1)-dimensional space-time. We
restrict ourselves to the case of homogeneous and isotropic brane which may
describe open, flat, or closed brane universe.

For a generally-covariant systems the Hamiltonian dynamics is encoded in a
system of constraints \cite{wdw}.  For a spherically symmetric space without
matter, and in any space-time dimensions, these constraints can be solved
explicitly classically as well as quantum mechanically, see
Ref.~\cite{kuchar}. This result can be understood noticing that in this case
gravity has only global degrees of freedom. The most convenient way to
parameterize these global degrees of freedom is to use the Schwarzschild-like
representation of the metric
\begin{equation}
\label{eq:metric}
ds^2=-F(t,r)dT^2+\frac{dR^2}{F(t,r)}+R^2d\Omega_3^2 \; ,
\end{equation}
where $T=T(t,r)$ and $R(t,r)$ are arbitrary functions of time and radial
coordinates $(t,r)$, while the function $F(t,r)$ has the form
\begin{equation}
\label{eq:F}
F(t,r)=k-\frac{l_{\rm Pl}^3M(t,r)}{R^2}-\Lambda R^2 \; ,
\end{equation}
where $k=0,\pm 1$ for the cases of flat, closed and open spatial sections,
respectively. 

In the Hamiltonian formalism the canonical variables describing the
bulk gravitational field are $(R,M;P_R,P_M)$. It turns out that
$T'=\partial T/\partial r$ is the momentum conjugate to $M$
\cite{kuchar}.  The conventional constraints of canonical formalism
reduce to the set of equations, $P_R=0$ and $M'=\partial M/\partial
r=0$.
One can see that if $M={\rm const}$, the metric (\ref{eq:metric}) coincides
with the metric of five-dimensional Schwarzschild-ant-deSitter black
hole of mass $M$.

The canonical constraint on the brane is
\begin{equation}
\label{constraint}
\hat H=\frac{3\hat R^2}{l_{\rm Pl}^3}\sigma\sqrt{|\hat F|}\cosh
\left\{\frac{l_{\rm Pl}^3\hat P_{\hat R}}{3\hat R^2}\right\}-
(\mu+\lambda\hat R^3)=0 \; ,
\end{equation}
where hat denotes the values of corresponding variables on the brane, e.g.
$\hat R=\left.R(t,r)\right|_{brane}$ and $\sigma = \pm 1$.  For the
geometrical meaning of the sign function $\sigma$ see Ref. \cite{shell1}.  At
the classical level $\sigma$ is integral of the motion, but the change of sign
is possible at the quantum level \cite{shell3,prd57}.  Note that
the Hamiltonian constraint Eq.~(\ref{constraint}) does not describe the
most general case (e.g.  the Schwarzschild parameter $M$ can be different on
both sides of the brane in general situation); rather, the $Z_2$ symmetry was
assumed following Ref.~\cite{rs}.  Positive (negative) sign of $\sigma$
corresponds to the positive (negative) brane tension in the case of classical
regime of Randall-Sundrum cosmology. For the discussion of general brane
Hamiltonian in the quantum case see Ref.~\cite{shell3,prd57}.

The equation of motion for $\hat R$ found from the
Hamiltonian~(\ref{constraint}) is
$d\hat R/d\tau=
\sigma\sqrt{|F|}\sinh\left(3l_{\rm Pl}^3 \hat P_R/\hat R^2\right)$,
which, upon substitution into (\ref{constraint}) gives  
\begin{equation}
\frac{(d\hat R/d\tau)^2}{\hat R^2}+\frac{k}{\hat R^2}=
\frac{l_{\rm Pl}^6(\mu+\lambda\hat R^3)^2}{9\hat R^6}+
\frac{l_{\rm Pl}^3M}{\hat R^4}+\Lambda\; .
\label{friedman}
\end{equation}
Being written in this form, the equation of motion of the brane resembles
closely the Friedmann equation \cite{br_Fr}, in which the density of matter on
the brane $\rho_m=\mu/{\hat R}^3$ enters quadratically at small ${\hat R}$,
the presence of non-zero bulk black hole mass $M$ results in the effective
``dark radiation'' contribution $\rho_{dr}=M/{\hat R}^4$ and the effective
cosmological constant on the brane is a certain combination of the bulk
cosmological constant and the brane tension
$\Lambda_{(3+1)}=l_{\rm Pl}^6\lambda^2/9+\Lambda$.
Note however that Eq.~(\ref{friedman}) is a ``square'' of true dynamical 
equation and important information encoded in $\sigma$ is lost. Therefore its
use can be inappropriate in some situations, especially in the quantum
regime.

{\it Quantum dynamics of the brane universe.}
In canonically quantized theory the Hamiltonian constraint (\ref{constraint})
is replaced by an operator equation on the wave function of the universe,
$\hat H\Psi=0$.
However, the quantization procedure in the coordinate representation would
result in the differential equation of infinite order. In addition, the
definition of operator
$$\cosh\left[l_{\rm Pl}^3 \hat P_R/3\hat R^2\right]=
\cosh\left[-i(l_{\rm Pl}^3/3\hat R^2)\partial/\partial \hat R)\right]$$ 
suffers from ambiguity related to the
operator ordering.

These problems can be solved if one makes canonical transformation
$v=\hat R^3;\ \ P_{v}
=\hat P_{\hat R}/(3\hat R^2)$ ,
which brings  the Hamiltonian $\hat H$ into the form
\begin{equation}
\label{hamilt}
\hat H=\frac{3v^{2/3}}{l_{\rm Pl}^3}\sigma
\sqrt{|\hat F|}\cosh\left\{l_{\rm Pl}^3
P_{v}\right\}-(\mu+\lambda v) \; .
\end{equation}
In the new variables, after quantization $P_v \rightarrow -i\partial/\partial
v$, the hyperbolic cosine which enters $\hat H$ becomes an operator of finite
shift along the imaginary axis,
$\exp\left(l_{\rm Pl}^3P_v\right)\Psi(v)=\Psi(v-il_{\rm Pl}^3)$.
Substituting this into $\hat H \Psi=0$ we find the following
finite-difference equation which determines the quantum dynamics of a
self-gravitating brane universe
\begin{eqnarray}
\label{eq:finite}
v^{2/3} F^{1/2}\left\{\Psi(v+il_{\rm Pl}^3)+
\Psi(v-il_{\rm Pl}^3)\right\}
\nonumber\\
-\frac{2}{3}\, l_{\rm Pl}^3(\mu+\lambda v)\Psi(v)&=&0.
\end{eqnarray}
Since the shift of the argument of the wave function is along imaginary 
axis, one has to consider the above equation in the complex plane, or, 
more precisely, on the  corresponding Riemanian  surface.  Indeed, the 
function  $F^{1/2}$ is a branching function on the complex plane.  The two 
branches,  $F^{1/2}=\pm\sqrt{F}$  correspond to the two possible choices 
of sigma. Therefore, if one finds the solutions of the above equation on 
the Riemann surface, the wave  function $\Psi$ is defined simultaneously in 
$\sigma=+1$ and $\sigma=-1$ domains.

In order to understand qualitatively the behavior of solutions 
of  Eq. (\ref{eq:finite}) we start with an analysis of the distances  
much larger than $l_{\rm Pl}$. In this limit
we can expand $\Psi(v\pm il_{\rm Pl}^3)$ in powers of the shift parameter, 
$\Psi(v\pm il_{\rm Pl}^3)\simeq \Psi(v)\pm il_{\rm Pl}^3\Psi'(v)-
\frac{1}{2}l_{\rm Pl}^6\Psi''(v)+ \dots$
In the first non-trivial order Eq.~(\ref{eq:finite}) reduces to
(we restrict ourselves to the case $\sigma=1$ here)
\begin{equation}
\label{eq:Schr}
\Psi''+
\frac{2}{l_{\rm Pl}^6}\left(1-\frac{l_{\rm Pl}^3 
(\mu+\lambda v)}{3v^{2/3}F^{1/2}}
\right)\Psi=0\; ,
\end{equation}
which is a Schroedinger-like  equation for particle motion in 
a potential
\begin{equation}
\label{potential}
U=1-\frac{l_{\rm Pl}^3 (\mu+\lambda v)}{3\left(
kv^{4/3}-2GMv^{2/3}+|\Lambda|v^2\right)^{1/2}} \; .
\end{equation}
For large $v=\hat R^3$ the
potential approaches a constant, $U\rightarrow 1- l_{\rm Pl}^3
\lambda/(3\sqrt{|\Lambda|})$. If
${l_{\rm Pl}^3 \lambda} > {3\sqrt{|\Lambda|}}$
the wave function behaves in the limit of large $R^3$ as a flat wave which
describes an expanding or contracting universe.

{\it Exactly solvable case of the flat universe.}
To make more detailed analysis of the quantum mechanics of the brane, e.g. to
study its spectrum, one needs to impose boundary conditions at the origin.  At
first sight the issue of boundary conditions at the Big Bang point $v=0$ looks
conceptually more simple for the brane universe.  Indeed, since the scale
factor of the universe is now just a position of the brane moving in the
external space (rather than purely gravitational degree of freedom), this is
just the question of boundary conditions on the wave function at the origin of
spherical coordinates .  However, in the region $v\sim l_{\rm Pl}^3$ 
one can not expand Eq. (\ref{eq:finite}) in powers of $l_{\rm Pl}$ and the
intuition based on Eq.  (\ref{eq:Schr}) is not applicable anymore. Instead,
one has to deal with the exact finite-difference equation (\ref{eq:finite}).
(We assume that the mini-superspace model based on the thin-wall 
approximation is still valid in the limit of small $v$.)

The finite-difference equations, and in particular the Eq. (\ref{eq:finite}),
possess a number of interesting general properties.  Being understood as an
infinite-order differential equations, they have to be supplemented with an
infinite set of boundary conditions. At the same time, starting from a single
particular solution $\Psi_0$ one can generate an infinite set of solutions
simply by multiplying $\Psi_0(z)$ by a function $C(z)$ which is periodic with
respect to the finite shift parameter (i.e. $C(z+il_{\rm Pl}^3)=C(z)$ in 
the case of Eq.~(\ref{eq:finite})). The appropriate methods of analysis of 
finite difference equations are discussed in 
Refs.~\cite{prd57,coulomb,hajicek}. 

In order to illustrate these methods it is convenient to consider the special
case when Eq.~(\ref{eq:finite}) is exactly solvable, namely the case of 
the flat universe $k=0$ and zero bulk Schwarzschild mass $M=0$. 
For this choise of parameters Eq.~(\ref{eq:finite}) takes the form
\begin{equation}
\label{eq:flat}
\Psi(v+il_{\rm Pl}^3)+\Psi(v-il_{\rm Pl}^3)
-\frac{2l_{\rm Pl}^3}
{3\sqrt{|\Lambda|}}\, \left(\lambda+\frac{\mu}{v}\right)\, \Psi(v)=0
\end{equation}
which coincides with the finite-difference analog of quantum-mechanical
problem of motion in Coulomb potential \cite{coulomb}.  A general solution of
Eq.~(\ref{eq:flat}) is given by (up to multiplication by an arbitrary 
$il_{\rm Pl}^3$ - periodic function)
$\Psi(S)=ve^{-\alpha v}{\cal F}(1-iv,1-\beta:2:1-e^{-2i\alpha})$,
where ${\cal F}$ is the hypergeometric function. Parameters 
$\alpha$ and $\beta$ 
are defined by relations $\cos\alpha=l_{\rm Pl}^3
\lambda/(3\sqrt{|\Lambda|})$ and $\beta\sin\alpha
=l_{\rm Pl}^3\mu/(3\sqrt{|\Lambda|})$. 

As it is usual in quantum mechanics, the single solution can be selected only
when the proper set of boundary conditions is chosen.   The correct boundary 
conditions can be determined from the requirement of vanishing of the 
probability flow
$J=i(\Psi^\dagger\hat H\Psi-\Psi\hat H\Psi^\dagger)$ at $v=0$. 
In the case of Eq.~(\ref{eq:flat}) this reduces to the set of 
conditions~\cite{hajicek}
$\Psi^{(2n)}(0)=0,\ n=0,1,....$

Similarly to the conventional quantum mechanics with the Coulomb
potential, there are bound states and continuous spectrum.  
Using the above boundary conditions  as well as appropriate conditions at
infinity, one can see that bound states exist when the
quantization condition
$$
\Lambda_{(3+1)}=\frac{\l_{\rm Pl}^6\lambda^2}{9}+\Lambda 
=- \frac{4\mu^2}{9n^2},\ ~~~n=1,2,...
$$
is satisfied. It relates the effective brane cosmological constant
$\Lambda_{(3+1)}$ to the matter density on the brane. In particular, the 
ground state of the universe corresponds to $n=1$.  
The wave functions of continuous spectrum  
($l_{\rm Pl}^3 \lambda>3\sqrt{|\Lambda|}$, 
which corresponds to the positive effective cosmological constant on the brane)
contain both the collapsing (in-going wave) and expanding (out-going)
branes. Thus, in the case of continuous spectrum, the wave function of the
universe corresponds to the so-called ``big bounce'' situation. One can
consider also transitions between the bound states and the states from
continuous spectrum (e.g. an expanding brane universe can result from 
the excitation of the ground state). However, the analysis of 
perturbations of the spherically symmetric system considered above goes 
beyond the mini-superspace approximation.

{\it Tunneling from the bound states.}
\begin{figure}
\resizebox{0.83\hsize}{!}{\includegraphics[angle=0]{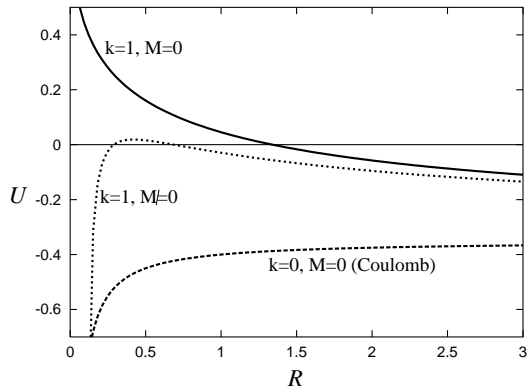}}
\caption{The potential $U$ (\ref{potential}) for different choices 
  of parameters. 
  The potential is singular at the gravitational radius of the
  bulk Schawrzscild-anti-deSitter black hole when $M\neq 0$. 
  }
\label{fig:potential}
\end{figure}
In order to study qualitatively the more general cases when the bulk
Schwarzschild mass in not zero let us come back to the analysis of the
truncated equation (\ref{eq:Schr}). The behavior of the potential
$U$ for the cases $k\not= 0$ and/or $M\not= 0$ is shown in
Fig. \ref{fig:potential}. One can see that if $l_{\rm Pl}^3M\ge
(\Lambda_{(3+1)})^{-1}$ there is a potential barrier, which separates
the regions of bound and unbound motion of the brane.  This means that
the spectrum of quantum states of the brane can contain, apart from
the discrete and continuous part, also ``resonances''. In this case
the expanding brane universe is the result of decay (or "tunneling")
of an almost stable state localized near the origin. The main
difference of the tunneling states considered here from the
(3+1)-dimensional ones is that in the $(3+1)$-dimensional case the
choice of the boundary conditions at $\hat R=0$ is ambiguous and the
existence of the "tunneling" state is, in fact, just
postulated~\cite{tunneling}.

{\it Discussion.}
In this paper we have constructed quantum cosmology of the brane universe and
have shown that it has several distinctive features.  In particular, one can
avoid the conceptual problems related to the interpretation of the wave
function of the universe. Indeed, in the brane-world setup one does not
quantize pure gravity, but rather deals with quantum mechanics of a matter
source (brane) moving through higher-dimensional space-time. The problem of
the choice of boundary conditions on the wave function of the universe is also
free from ambiguities: one simply has to impose the usual quantum mechanical
conditions on the wave function at the origin of coordinates.  This allows for
the detailed analysis of bound states, continuous spectrum and tunneling
states, where creation of the universe from ``nothing'' can be interpreted as
a decay of a bound state resonance.

When gravitational self-interaction of the brane universe is
important, as, for example in the setup of Randall-Sundrum cosmology studied
here, one has to correctly account for the bulk-brane interaction not only
classically, but also on quantum level. As a result, the classical
brane Hamiltonian constraint (\ref{constraint}) becomes after quantization a
finite-difference equation (\ref{eq:finite}).

Although the appearance of finite-difference equations is a novel feature of
the quantum brane cosmology, the analysis of the boundary
conditions and of the wave functions of discrete and continuous spectra can be
carried in a way similar to the one used in conventional quantum mechanics.
From the point of view of quantization of gravitating systems the appearance
of a non-local equation (with non-locality at the Plank scale) is natural to
expect.  Such equations appear in several other models (see e.g.
\cite{thooft,rujsenaars,snyder,nekrasov}). It implies a deformation of the
Lorentz symmetry and generalized uncertainty principle
\cite{jacobson,amelino}.
\vskip-0.5cm




\begin{thebibliography}{99}

\bibitem{mini} Y.~B.~Zeldovich and A.~A.~Starobinsky,
Sov.\ Astron.\ Lett.\  {\bf 10}, 135 (1984);
A.~D.~Linde,
Sov.\ Phys.\ JETP {\bf 60}, 211 (1984)
[Zh.\ Eksp.\ Teor.\ Fiz.\  {\bf 87}, 369 (1984)];
A.~Vilenkin,
Phys.\ Rev.\ D {\bf 30}, 509 (1984).
V.~A.~Rubakov,
Phys.\ Lett.\ B {\bf 148}, 280 (1984).

\bibitem{tunneling}
A.~Vilenkin,
Phys.\ Lett.\ B {\bf 117}, 25 (1982) and
Phys.\ Rev.\ D {\bf 27}, 2848 (1983).

\bibitem{Linde:2004nz}
A.~Linde,
JCAP {\bf 0410}, 004 (2004).

\bibitem{string}
A.~Giveon, E.~Rabinovici and A.~Sever,
Fortsch.\ Phys.\  {\bf 51}, 805 (2003).

\bibitem{wdw}
B.~S.~Dewitt,
Phys.\ Rev.\  {\bf 160}, 1113 (1967).

\bibitem{qg}
J.~B.~Hartle and S.~W.~Hawking,
Phys.\ Rev.\ D {\bf 28}, 2960 (1983).

\bibitem{jan}
J.~Ambjorn, J.~Jurkiewicz and R.~Loll,
hep-th/0411152.



\bibitem{vilenkin_rev} 
A.~Vilenkin,
gr-qc/020461.

\bibitem{Rubakov:1983bb}
V.~A.~Rubakov and M.~E.~Shaposhnikov,
Phys.\ Lett.\ B {\bf 125}, 136 (1983).

\bibitem{rs} 
L.~Randall and R.~Sundrum,
Phys.\ Rev.\ Lett.\  {\bf 83}, 4690 (1999).


\bibitem{shell1}
V.~A.~Berezin, V.~A.~Kuzmin and I.~I.~Tkachev,
Phys.\ Lett.\ B {\bf 120}, 91 (1983) and
Phys.\ Rev.\ D {\bf 36}, 2919 (1987).

\bibitem{shell1a}
A.~Vilenkin,
Phys.\ Lett.\ B {\bf 133}, 177 (1983);
J.~Ipser and P.~Sikivie,
Phys.\ Rev.\ D {\bf 30}, 712 (1984);
S.~K.~Blau, E.~I.~Guendelman and A.~H.~Guth,
Phys.\ Rev.\ D {\bf 35}, 1747 (1987);
A.~Aurilia, R.~S.~Kissack, R.~Mann and E.~Spallucci,
Phys.\ Rev.\ D {\bf 35}, 2961 (1987).

\bibitem{shell2}
V.~A.~Berezin, N.~G.~Kozimirov, V.~A.~Kuzmin and I.~I.~Tkachev,
Phys.\ Lett.\ B {\bf 212}, 415 (1988).

\bibitem{shell3}
W.~Fischler, D.~Morgan and J.~Polchinski,
Phys.\ Rev.\ D {\bf 42}, 4042 (1990).


\bibitem{prd57}  
V.~A.~Berezin, A.~M.~Boyarsky and A.~Y.~Neronov,
Phys.\ Rev.\ D {\bf 57}, 1118 (1998) and
Phys.\ Lett.\ B {\bf 455}, 109 (1999);
A.~Y.~Neronov,
Phys.\ Rev.\ D {\bf 59}, 044023 (1999).

\bibitem{Bojowald}
M.~Bojowald,
Phys.\ Rev.\ Lett.\  {\bf 86}, 5227 (2001);
Phys.\ Rev.\ Lett.\  {\bf 87}, 121301 (2001) and
Phys.\ Rev.\ Lett.\  {\bf 89}, 261301 (2002).

\bibitem{kuchar}
K.~V.~Kuchar,
Phys.\ Rev.\ D {\bf 50}, 3961 (1994).


\bibitem{br_Fr}
J.~M.~Cline, C.~Grojean and G.~Servant,
Phys.\ Rev.\ Lett.\  {\bf 83}, 4245 (1999);
P.~Binetruy, C.~Deffayet and D.~Langlois,
Nucl.\ Phys.\ B {\bf 565}, 269 (2000);
P.~Brax, C.~van de Bruck and A.~C.~Davis,
Rept.\ Prog.\ Phys.\  {\bf 67}, 2183 (2004);
R.~Maartens,
Phys.\ Rev.\ D {\bf 62}, 084023 (2000);
A.~Neronov,
gr-qc/0101060;
A.~Neronov,
JHEP {\bf 0111}, 007 (2001).



\bibitem{coulomb} 
 V.~G.~Kadyshevsky, R.~M.~Mir-Kasimov and M.~Freeman,
Yad.\ Fiz.\  {\bf 9}, 646 (1969);
V.~A.~Berezin,
Phys.\ Rev.\ D {\bf 55}, 2139 (1997).



\bibitem{hajicek} 
P.~Hajicek,
Commun.\ Math.\ Phys.\  {\bf 150}, 545 (1992).

\bibitem{rujsenaars}
S.~N.~M.~Ruijsenaars and H.~Schneider,
Annals Phys.\  {\bf 170}, 370 (1986).

\bibitem{snyder}
H.~S.~Snyder,
Phys.\ Rev.\  {\bf 71}, 38 (1947).


\bibitem{nekrasov}
A.~Gorsky and N.~Nekrasov,
Nucl.\ Phys.\ B {\bf 436}, 582 (1995).

\bibitem{thooft} 
G.~'t Hooft,
Class.\ Quant.\ Grav.\  {\bf 13}, 1023 (1996).



\bibitem{jacobson}
T.~Jacobson, S.~Liberati and D.~Mattingly,
gr-qc/0404067.

\bibitem{amelino}
G.~Amelino-Camelia,
gr-qc/0309054.

\end{thebibliography}
\end{document}